# Molding Photon Emission with Hybrid Plasmon-Emitter Coupled Metasurfaces

*Yinhui Kan[1] and Sergey I. Bozhevolnyi[2]\**

[1]College of Astronautics, Nanjing University of Aeronautics and Astronautics, Nanjing 210016, China

[2]Center for Nano Optics, University of Southern Denmark, DK-5230 Odense M, Denmark

**Abstract:** Directional emission of photons with designed polarizations and orbital angular momenta is crucial for exploiting full potential of quantum emitters (QEs) within quantum information technologies. Capitalizing on the concept of hybrid plasmon-QE coupled metasurfaces, we develop a holography-based design approach allowing one to construct surface nanostructures for outcoupling QE-excited circularly diverging surface plasmon polaritons (SPPs) into well-collimated beams of photons with desirable polarization characteristics propagating along given directions. Using the well-established simulation framework, we demonstrate the efficiency and versatility of the developed approach by analyzing different hybrid SPP-QE coupled metasurfaces designed for generation of linearly, radially and circularly polarized photons propagating in various off-axis directions. Our work enables the design of single-photon sources with radiation channels that have distinct directional and polarization characteristics, extending thereby possibilities for designing complex photonic systems for quantum information processing. Moreover, we believe that the developed scattering holography approach is generally useful when employing surface electromagnetic excitations, including SPP modes, as reference waves for signal wave reconstruction.

Many efforts have been dedicated over recent years to improving the performance of quantum emitters (QEs) that constitute one the key enabling technologies for quantum communication and information systems[1-5]. Various of micro/nano structures, including nanoantennas and nanocavities, have been designed in recent years to enhance the photon emission from QEs by making use of the Purcell effect[6-11] via engineering their immediate dielectric environment[12-22]. Although very large (even in thousands fold) enhancement of brightness has been reported with different nanostructures, the polarization, direction, and wavefront of the photon emission are still rarely discussed[16-22]. Very recently, the design route for optical metasurfaces has been introduced to efficiently mold the single-photon emission from QEs with hybrid plasmon-QE coupled configurations, consisting of dielectric nanoridges surrounding QEs on metal substrates, that appear most attractive from the viewpoint of combining high quantum yield and collection efficiency[23-28]. A common feature of these configurations (meta-atoms) is *efficient* and *nonradiative* coupling of a QE to surface plasmon-polariton (SPP) modes that are subsequently outcoupled by a metasurfaces structure into free propagating waves[23]. Different designs of the outcoupling metasurface resulted in generation of circularly (with different orbital angular momenta)[23, 24] or radially[27] polarized single photons. Using displaced circular ridges, directional off-normal photon streaming for different, up to 20°, angles has also been experimentally realized[28]. At the same time, serious deficiencies in generated wavefronts were observed both in simulations and experiments, especially when producing off-normal photon streaming[28], calling for the development of the general design approach that would ensure the realization of well-collimated beams of photons with desirable polarizations (e.g., linear, circular, and radial) propagating in given (off-axis) directions.

Holography is an approach enabling the recording and reconstruction of arbitrary wavefronts with the help of various types (amplitude, phase, transmission, reflection, *etc*.) of holograms[29-34]. The problem of generating complicated wavefronts with arbitrarily designed polarizations distributions is of great importance in photonics, in general, and in the field of optical metasurfaces, in particular[35-40]. Very recently, efficient approaches using local control of the phase and polarization of transmitted/reflected light have been developed for designing optical metasurfaces that perform rather complicated transformations of wavefronts and polarizations in transmission[41] and reflection[42] configurations. Although the main principles of holography and its many modalities are well known and established for free propagating (object and reference) optical beams, the problem of reconstructing beams with arbitrary wavefronts and polarizations when using (QE-excited) circularly diverging SPP fields has not been considered. In this work, we introduce a novel modification, vectorial scattering (computer-generated) holography, for the purpose of designing hybrid SPP-QE coupled metasurfaces suitable for generation of well-collimated beams of single photons with desirable polarization characteristics propagating along given directions. With this holography-based approach at hand, we design metasurfaces for the generation of linearly, radially or circularly polarized photons propagating in various off-axis directions and demonstrate the efficiency and versatility of our approach using simulations conducted within the well-established numerical framework[23, 24, 27, 28].

The basic configuration under consideration, representing the hybrid SPP-QE coupled metasurface, consists of dielectric nanoridges (forming a metasurface) formed atop a metal substrate covered with a dielectric spacer (Figure 1). We have previously demonstrated that such a meta-atom can be designed to very efficiently (up to 90%) convert QE-excitations into free propagating and objective-collected photons[27]. In the simulations presented here, a QE considered is a nitrogen vacancy (NV) center in nano-diamond (ND), whose emission peak wavelength is near 670 nm (when illuminated by a 532 nm laser beam) and whose radiative transition dipole is oriented perpendicular to the sample surface[23, 24, 27, 43]. The QE-excited SPPs propagate along a silica (15 nm thickness) covered silver film, circularly diverging from the QE in a cylindrically symmetric fashion. Nanoridges made of hydrogen silsesquioxane (HSQ, 180 nm thickness) with the refractive index of 1.41[28, 44] are positioned atop the silica spacer layer (protecting a silver film from oxidization). The pattern of nanoridges encircling the QE is designed in accordance with the holography-based approach developed in this work that enables the SPP outcoupling into a single-photon stream carrying a specific (linear, radial, or circular) polarization and propagating along a given direction.

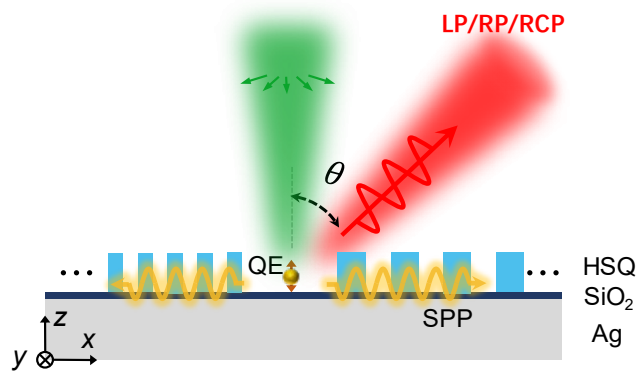

**Figure 1.** Schematic representation of hybrid SPP-QE coupled metasurfaces considered in this work for generation of well-collimated beams of single photons with desirable polarization characteristics, such as linear (LP), radial (RP) or right circular (RCP) polarizations, propagating along given directions.

To design the nanoridge pattern of a hybrid SPP-QE coupled metasurface, we develop a modified vectorial scattering holography approach (Supporting Information, Sections 1 and 2) with a cylindrically diverging SPP wave, $\boldsymbol{E}_{spp}$, used to reconstruct a signal wave, $\boldsymbol{E}_s$. In brief, a hologram in our approach is formed by *isotropic non-interacting dipolar nano-scatterers* (such as spherical nanoparticles), whose volumes and thereby scalar polarizabilities are proportional to the intensity of an interference pattern generated by a reference and signal waves. Discretizing the intensity pattern and fusing neighbor nanoparticles results in a pattern of constant-height nanoridges (of varying width) that can conveniently be fabricated with the electron-beam lithography (EBL). Importantly and *contrary* to the conventional holography approaches[29-34], we employ an *artificial* reference wave for calculating the interference pattern and designing thereby the nanoridge pattern. This artificial reference wave is taken in the form of a *radially increasing* SPP wave that can be expressed in the hologram plane as follows: $\boldsymbol{E}_{a,spp} = \boldsymbol{E}_{spp}^0 \cdot \sqrt{r_{xy}} \cdot \exp(\alpha r_{xy}) \exp(-ik_{spp}r_{xy})$, where $\boldsymbol{E}_{spp}^0$ is the SPP vectorial amplitude, $r_{xy}$ is the length of radius vector, $k_{spp} = (2\pi/\lambda)N_{spp}$, with $N_{spp}$ being the SPP effective index at the operation wavelength $\lambda$, and $\alpha = 1/2L_{spp}$, with $L_{spp}$ being the SPP propagation length. The nanoridge pattern is then calculated by considering interference between the above artificial SPP wave and the signal wave (taken also in the hologram plane): $\boldsymbol{E}_s = \boldsymbol{E}_s^0 \exp(-i\boldsymbol{k}_s^{xy} \cdot \boldsymbol{r}_{xy})$, where $\boldsymbol{E}_s^0$ and $\boldsymbol{k}_s^{xy}$ are the amplitude and wavevector projection (on the hologram plane) of the signal wave, respectively (both quantities are assumed to be varying in the hologram plane). When the resulting nanoridge pattern is illuminated by the actual (QE-excited) SPP wave: $\boldsymbol{E}_{spp} = \boldsymbol{E}_{spp}^0 \cdot (1/\sqrt{r_{xy}}) \cdot \exp(-\alpha r_{xy}) \exp(-ik_{spp}r_{xy})$, the scattered field in the (*x*, *y*)-plane near the hologram contains the corresponding (to the reconstructed signal wave) term in the following form (Supporting Information, Section 2):

$$E_{sc} \sim \ldots + (E_{spp}^{0*} \cdot E_s^0) E_{spp}^0 \exp(-i k_s^{xy} \cdot r_{xy}) \quad . \tag{1}$$

Although the problem of reconstructing the signal wave polarization remains when one uses *isotropic* dipolar nano-scatterers (Supporting Information, Section 1), the spatial phase and amplitude distribution of the signal wave (encoded in the signal amplitude and wavevector projection spatial distributions) are expected to be reconstructed, ensuring the propagation of a specified beam in a specified direction. Note that the SPP depletion during the reconstruction, occurring due to the SPP scattering out of the surface plane by surface nanoparticles, can, in principle, also be accounted for in a similar manner by simply adding the SPP attenuation by scattering to the SPP attenuation by absorption: $\alpha^* = \alpha + \alpha_{scat}$. In the following we illustrate this approach for designing nanoridge patterns of hybrid SPP-QE coupled metasurfaces that generate differently polarized and propagating single-photon beams. In view of reconstruction problems associated with the polarization crosstalk (Supporting Information, Section 1), we limit calculating the interference (and thereby designing the nanoridge pattern) to considering only in-plane (*x*, *y*)-components of reference and signal waves. For reconstruction, we employ three-dimensional (3D) finite difference time domain (FDTD) simulations using the well-established (theoretically and experimentally) numerical framework[23, 24, 27, 28].

First, we design metasurfaces for generating linearly polarized, along the *x*-axis (LPX), Gaussian beams of different radii $w_0$ propagating normal to the metasurface plane ($k_s^{xy} = 0$), so that the only non-zero signal field component is the *x*-component: $E_{sx} \sim \exp(-r_{xy}^2/w_0^2)$. The considered beam radii are conveniently related to the SPP propagation length (8 μm), for example $w_0 = L_{spp}/\sqrt{2}$ (Figure 2), that was calculated using the average filling ratio (0.39) of HSQ found from the designed nanoridge pattern. As expected, the nanoridges vanish near the *y*-axis [Figure 2(a)], where the polarizations of the reference and SPP wave are orthogonal to each other, resulting in the reconstructed LPX beams becoming elliptical (less confined along the *y*-axis) in the cross-section [Figure 2(b, c)]. Importantly, the power carried by the LPX beams is significantly larger than that of LPY (four-lobed) beams, with the total field along the normal to the hologram center being purely *x*-polarized. Finally, the beam divergence angle $\vartheta$ is found decreasing when the Gaussian signal beam radius increases, closely following the theoretically expected dependence: $\vartheta = \lambda/\pi w_0$ [Figure 2(d)], because wider signal beams result in larger intensity interference patterns and thereby hologram areas (Supporting Information, Figure S5).

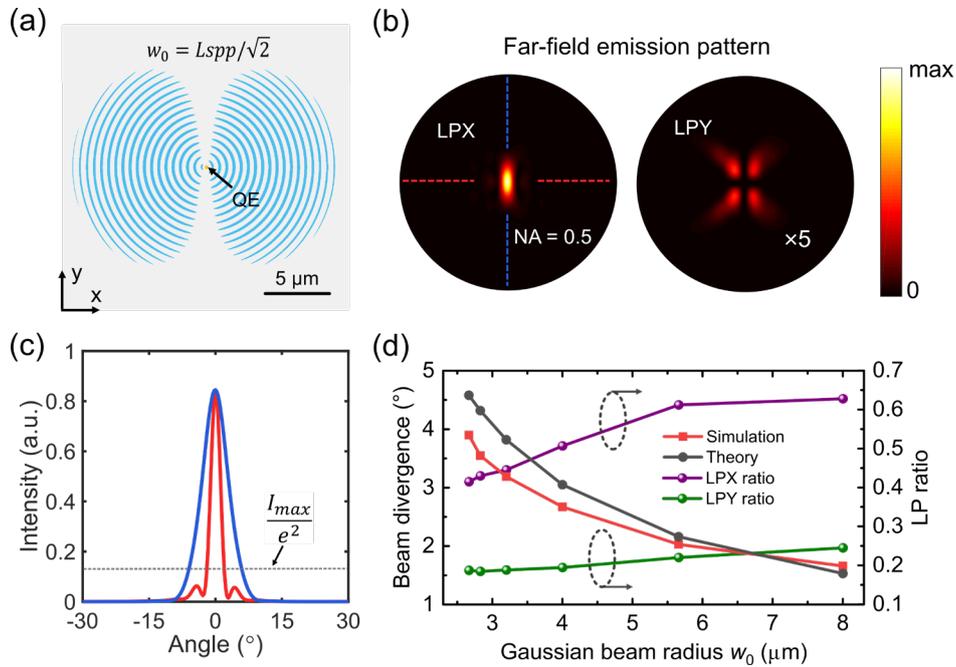

**Figure 2.** Directional linearly polarized photon emission from hybrid SPP-QE coupled metasurfaces. (a) Top view of the nanoridge pattern, (b) far-field emission patterns (NA = 0.5), and (c) far-field
7

angular intensity cross-sections for the LPX beam reconstructing the Gaussian signal beam with the radius $w_0 = L_{spp}/\sqrt{2}$. (d) LPX beam divergence and relative (to the total emitted power) powers of LPX and LPY beams as a function of the signal beam radius $w_0$.

To realize the off-axis LPX photon emission, we consider the *x*-polarized signal field in the following form: $E_{sx} \sim \exp(-r_s^2/w_0^2) \exp(ik_0 \sin\theta\, x)$, where $k_0 = 2\pi/\lambda$, $\theta$ is the emission angle in the (*x*, *z*)-plane with respect to the surface normal and $r_s^2 = (x\cos\theta)^2 + y^2$. Calculating the corresponding interference patterns, one notices that these and, consequently, nanoridge patterns become progressively more asymmetric for larger emission angles (Supporting Information, Figure S6). At the same time, it is seen that the reconstruction remains faithful, reproducing well-confined LPX beams propagating in the designed direction even for rather large, up to $\theta = 60°$, angles (Figure 3), which is very important from the viewpoint of potential applications in quantum technologies.

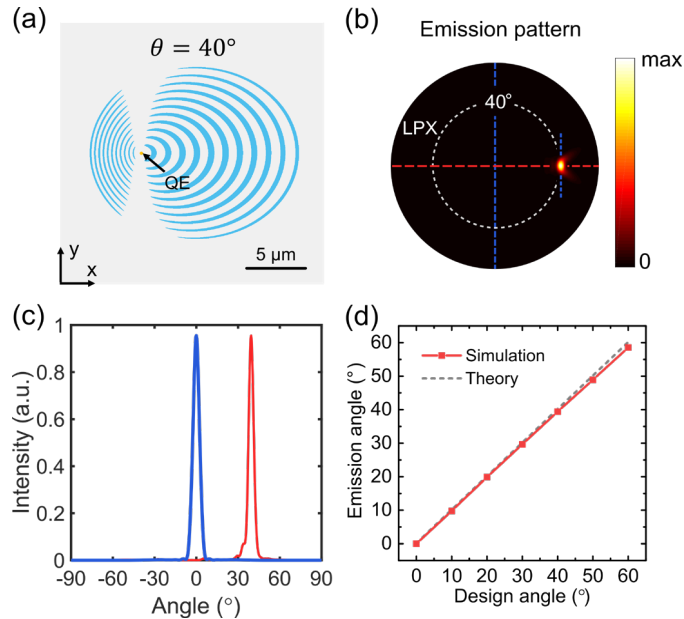

**Figure 3.** Off-axis LPX photon emission from hybrid SPP-QE coupled metasurfaces. (a) Top view of the nanoridge pattern, (b) far-field LPX emission pattern, and (c) far-field angular intensity cross-

sections for the LPX beam reconstructing the Gaussian signal beam propagating at the design angle $\theta = 40°$. (d) The LPX emission angles obtained for different designed metasurfaces along with the designed angle used.

Generation of radially polarized (RP) photon beams propagating normal to the surface is relatively straightforward with a simple bullseye nanoridge pattern[27, 45,46] because of the same symmetry for in-plane $(x, y)$-components of both QE-excited radially diverging SPP waves and RP beams. The situation becomes more complicated for off-axis RP emission: the Doppler-like design of displaced circular ridges does not work well, failing to reconstruct the doughnut shape even for small ($\theta < 10°$) angles[28.] Our unified design strategy suggests using the in-plane RP field distribution (signal wave) for calculating the interference pattern in the following form: $\boldsymbol{E}_s = (E_{sx}; E_{sy}) \sim (r_s/w_0) \exp(-r_s^2/w_0^2) \exp(ik_0 \sin\theta\, x) \cdot (\cos\theta \cos\varphi\,;\, \sin\varphi)$, where $\varphi$ is the polar angle: $\tan\varphi = y/x$. It is seen that for relatively large angles, such as $\theta = 15°$, the doughnut intensity distribution is well reconstructed when using the designed nanoridge pattern [Figure 4(a)], although some distortion is visible [Figure 4(b)]. The latter is believed to be due to relatively stronger forward SPP scattering (with respect to the backward one) by nanoridge arrays[27], a circumstance that results in changing the field balance along the RP beam circumference. Importantly, the RP field decomposition into circularly polarized fields reveals opposite orbital angular momenta or topological charges ($\pm 1$) in their phases [Figure 4(c, d)] as expected[24]. For larger angles the doughnut shape is progressively more distorted, but the emission angles remain very close to the designed values [Supporting Information, Figure S7(a)].

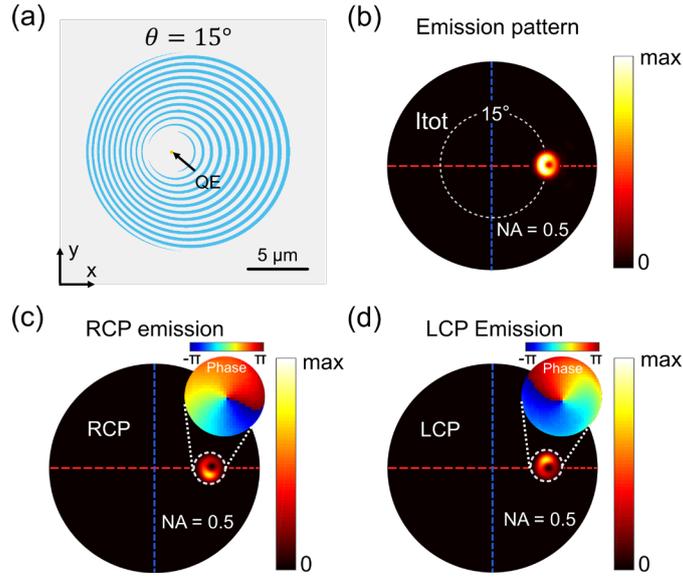

**Figure 4.** Off-axis RP photon emission from hybrid SPP-QE coupled metasurfaces for the emission angle $\theta = 15°$. (a) Top view of the nanoridge pattern, (b) far-field RP emission pattern, (c) RCP and (d) LCP emission patterns, with insets showing the corresponding phase distributions.

Generation of purely circularly right or left polarized (RCP or LCP) photon beams is impossible with simple spiral nanoridge patterns that would always generate RCP and LCP beams carrying orbital angular momenta (topological charges) different by 2[24]. We can still use our approach targeting the generation of, for example, RCP beam and thus using the following signal wave for calculating the interference pattern: $\boldsymbol{E}_s = (E_{sx}; E_{sy}) \sim \exp(ik_0 \sin\theta\, x) \cdot (\cos\theta; i)$. It is seen that for relatively large angles, such as $\theta = 15°$, the well-defined RCP beam and the doughnut-shaped (vortex) LCP beam, carrying the orbital angular momentum with the topological charge of 2, are both well reconstructed (Figure 5), i.e., showing the expected intensity distribution[24]. Small distortion in the LCP intensity distribution is seen [Figure 5(d)] similarly to that observed for the RP case above and being also related to the asymmetry in the off-axis SPP scattering by nanoridge arrays. Note, that the resulting emission pattern [Figure 5(b)] consists of spatially separated RCP [Figure 5(c)] and LCP [Figure 5(d)] emission channels, representing thereby spatially separated entangled emission channels, a feature that is very important from the viewpoint of potential applications in quantum technologies [24]. For larger angles the doughnut LCP beam shape is progressively more distorted, but the RCP beam remains well shaped and confined, with the emission angles being very close to the designed values [Supporting Information, Figure S7(b)].

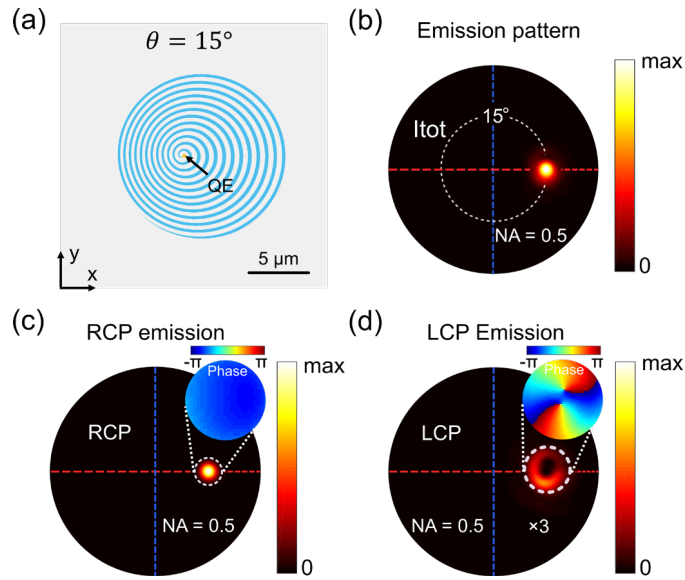

**Figure 5.** Off-axis circularly polarized photon emission from hybrid SPP-QE coupled metasurfaces for the emission angle $\theta = 15°$. (a) Top view of the nanoridge pattern, (b) far-field emission pattern, (c) RCP and (d) LCP emission patterns, with insets showing the corresponding phase distributions.

In summary, we have developed a holography-based design approach for constructing nanoridge patterns of hybrid SPP-QE coupled metasurfaces enabling generation of well-collimated single-photon beams with desirable wavefront and polarization characteristics propagating along given directions. Conducting 3D-FDTD simulations using the well-established (theoretically and experimentally) numerical framework[23, 24, 27, 28], we have demonstrated the efficiency and versatility of the developed approach by analyzing different hybrid SPP-QE coupled metasurfaces designed for generation of LP, RP and CP polarized photon beams propagating in various off-axis directions. We would like to note that the presented approach can further be developed and generalized by using properly designed *anisotropic* (instead of isotropic) nano-scatterers (Supporting Information, Section 1). This generalization should enable increasing the generation efficiency up to the level, ~ 90%[27] expected for the RP generation (when there is a perfect match between in-plane field components of the reference SPP and signal RP waves). Overall, our work opens fascinating perspectives for designing single-photon sources with radiation channels that exhibit diverse (including vectorial with spin and orbital angular momenta) wavefronts and polarization characteristics, extending thereby possibilities for designing complex photonic systems for quantum information processing. Furthermore, the developed scattering holography approach allows one to achieve faithful reconstruction of complex signal wavefronts in configurations that involve surface electromagnetic excitations, including SPP modes, as reference waves by accounting for inevitable (due to divergence, absorption and scattering losses) spatial variations of reference wave magnitude (Supporting Information, Section 2).

ASSOCIATED CONTENT

**Supporting Information**.

Section S1: Vectorial scattering holography: fundamentals. Section S2: Vectorial scattering holography: using surface plasmon polaritons as reference waves. Section S3: Numerical simulation.

AUTHOR INFORMATION

**Corresponding Author**

*(S.I.B.) E-mail: seib@mci.sdu.dk

ACKNOWLEDGEMENTS

Y.H.K. acknowledge the support from National Natural Science Foundation of China (Grant No. 62105150), Natural Science Foundation of Jiangsu Province (BK20210289), Fundamental Research Fund for Central Universities (NS2021064). The authors gratefully acknowledge the Villum Kann Rasmussen Foundation (Award in Technical and Natural Sciences 2019).

**Supporting Information**

Supporting Information Available:

The fundamentals of the vectorial scattering holography; The detailed description of the proposed holography-based design approach; The details of numerical simulations for the different kinds of hybrid SPP-QE coupled metasurfaces design.

This material is available free of charge via the Internet at http://pubs.acs.org

# Supporting Information

# Molding Photon Emission with Hybrid Plasmon-Emitter Coupled Metasurfaces

Yinhui Kan[1] and Sergey I. Bozhevolnyi[2]*

[1] College of Astronautics, Nanjing University of Aeronautics and Astronautics, Nanjing 210016, China

[2] Center for Nano Optics, University of Southern Denmark, DK-5230 Odense M, Denmark

## 1. Vectorial scattering holography: fundamentals

In general, holography is an approach enabling the recording and reconstruction of arbitrary wavefronts with the help of various types (amplitude, phase, transmission, reflection etc.) of holograms [S1]. One can also design and fabricate holograms for reconstruction of desirable wavefronts by computing interference patterns and using various fabrication techniques. This approach is usually called computer-generated holography [S1]. Although the main principles of holography and its many modalities are well known and established, we should introduce a novel modification, vectorial scattering (computer-generated) holography, for the purpose of designing hybrid plasmon-QE coupled metasurfaces suitable for generation of well-collimated beams of photons with desirable polarization characteristics propagating along given directions.

We start our consideration with introducing scalar scattering holography, in which all fields are scalar. As the first step, we consider the intensity interference pattern produced by electrical fields of a signal, $E_s$, and reference, $E_r$, waves in the hologram $xy$-plane, $z = 0$ [Figure S1(a)]:

$$I(x,y) = |E_r + E_s|^2 = \left|E_r^0 \exp(-i\mathbf{k}_r^{xy} \cdot \mathbf{r}_{xy}) + E_s^0 \exp(-i\mathbf{k}_s^{xy} \cdot \mathbf{r}_{xy})\right|^2 =$$
$$= I_r^0 + I_s^0 + E_r^0 E_s^{0*} \exp[-i(\mathbf{k}_r^{xy} - \mathbf{k}_s^{xy}) \cdot \mathbf{r}_{xy}] + E_r^{0*} E_s^0 \exp[-i(\mathbf{k}_s^{xy} - \mathbf{k}_r^{xy}) \cdot \mathbf{r}_{xy}], \quad (S1)$$

where $E_s^0$ and $E_r^0$ are the complex electric field amplitudes, $\mathbf{k}_s^{xy}$ and $\mathbf{k}_r^{xy}$ – the wavevector projections on the hologram $xy$-plane of the signal and reference waves, respectively (all quantities are assumed to be varying in the hologram plane), $\mathbf{r}_{xy}$ is the radius vector. Let us consider a hologram, which is formed by isotropic non-interacting dipolar nano-scatterers (such as spherical nanoparticles), whose volumes and thereby scalar polarizabilities are proportional to the intensity of the interference pattern. The field scattered by such a hologram, when illuminated with the reference wave, can then be expressed in the immediate vicinity of nanoparticles located at discrete set of points $(x_i, y_i)$ as follows:

$$E_{sc}(x_i, y_i) \sim (IE_r)_{x_i,y_i} = \{(I_r^0 + I_s^0)E_r^0\exp(-i\boldsymbol{k}_r^{xy} \cdot \boldsymbol{r}_{xy}) + E_r^0 E_s^{0*} E_r^0 \exp[-i(2\boldsymbol{k}_r^{xy} - \boldsymbol{k}_s^{xy}) \cdot \boldsymbol{r}_{xy}] +$$
$$+ |E_r^0|^2 E_s^0 \exp(-i\boldsymbol{k}_s^{xy} \cdot \boldsymbol{r}_{xy})\}_{x_i,y_i} \, . \tag{S2}$$

Assuming that (<u>non-interacting</u>) nanoparticles are densely packed at subwavelength distances, one arrives at the usual result of any holographic approach [S1] that the field produced by a hologram consists of three terms: an addition to the incident reference wave (first term), the phase-conjugated signal wave (second term) and the reconstructed signal wave (third term). These terms can also be thought as representing three main diffraction orders, i.e., the $0^{th}$ and $\pm 1^{st}$ diffraction orders [Figure S1(b)]. Faithful reconstruction of the signal wave requires <u>non-interacting isotropic</u> nanoparticles and the reference wave being close to a homogeneous plane wave (the latter being the usual requirement [S1]).

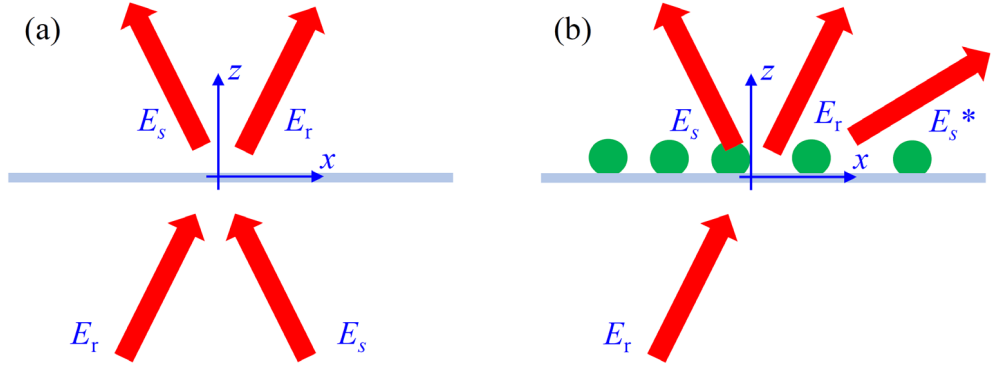

**Figure S1.** Schematic of vectorial scattering holography, illustrating (a) the determination of the intensity interference pattern produced by a signal, $E_s$, and reference, $E_r$, waves in the hologram $xy$-plane, and (b) the reconstruction of the signal wave upon interaction of the reference wave with the arrays of surface nanoparticles located at intensity maxima of the interference pattern.

The situation is significantly more complicated when considering vector reference and signal fields. In this case, the analogue of Eq. (S2) becomes (omitting superfluous indexes):

$$\boldsymbol{E}_{sc} \sim I\boldsymbol{E}_r = (I_r^0 + I_s^0)\boldsymbol{E}_r^0\exp(-i\boldsymbol{k}_r^{xy} \cdot \boldsymbol{r}_{xy}) +$$
$$+ (\boldsymbol{E}_r^0 \cdot \boldsymbol{E}_s^{0*})\boldsymbol{E}_r^0 \exp[-i(2\boldsymbol{k}_r^{xy} - \boldsymbol{k}_s^{xy}) \cdot \boldsymbol{r}_{xy}] + (\boldsymbol{E}_r^{0*} \cdot \boldsymbol{E}_s^0)\boldsymbol{E}_r^0 \exp(-i\boldsymbol{k}_s^{xy} \cdot \boldsymbol{r}_{xy}) \, . \tag{S3}$$

It is immediately seen that, while the waves representing the incident, phase-conjugated signal and signal terms have the corresponding propagation directions, their polarization properties, *in the near* *field*, are identical to that of the reference wave. In a particular case of both signal and reference waves being TE-polarized (i.e., perpendicular to the plane of incidence) and propagating at close to normal incidence (i.e., within the paraxial approximation), the above expression can be reduced to that obtained for scalar waves [Eq. (S2)]. In this case, a faithful reconstruction of the signal wave is thereby expected. In a general case of differently polarized signal and reference waves, the most serious problem would appear if the electric fields of signal and conjugate reference waves would be orthogonal to each other, at least in some areas of the interference plane, thereby eliminating the corresponding interference term and loosing the phase information. It should be noted that, *in the far field*, the reconstructed signal wave (last term) does not necessarily exhibit the same polarization as that of the reference wave because of far-field interference effects.

The problem of generating complicated wavefronts with arbitrarily designed polarizations distributions is of great importance in photonics, in general, and in the field of optical metasurfaces, in particular [S2]. Given the possibility of locally control the phase and polarization of transmitted or reflected light, efficient approaches have been developed for designing optical metasurfaces operating in transmission [S3] or reflection [S4] and performing rather complicated transformations of wavefronts and polarizations. In principle, one can also extend the design parameters within the considered scattering holography, e.g., by introducing ellipsoidal nanoparticles with spatially varying ellipticity and orientation, to enable spatially varying polarization transformation from the local reference field polarization to the signal one. It seems reasonable to expect that, by using properly designed anisotropic nanoparticles, one should be able of faithfully reconstructing signal waves having arbitrarily distributed amplitudes, phases and polarizations. Developing this generalization is a rather challenging issue that goes however beyond the framework of the current work. In our present paper, we concentrate on considering polarization and wavefront transformations that are feasible within the current holography concept, in which closely spaced isotropic nanoparticles are merged into nano-ridges forming a pattern representing a digitized intensity interference pattern.

Concluding fundamentals of vectorial scattering holography, let us examine the consequences of making up the intensity interference pattern from differently polarized reference and signal waves. To simplify our analysis, we consider both propagating waves in the paraxial approximation, so that only in-plane field components are taken into account. In this case, the *x*- and *y*-components of the signal and reference fields can be treated separately, resulting in the corresponding intensity interference patterns, $I_x$ and $I_y$, that can be expressed similarly to that obtained for the scalar case [see Eq. (S1)] and should be added together to obtain the resulting interference pattern. Since the scattering hologram is made of spherical nanoparticles, there is no cross-polarized scattering, and the *x*-component of the scattered field can be expressed as follows:

$$E_{sx} \sim IE_{rx} = (I_r^0 + I_s^0)E_{rx}^0 \exp(-i\mathbf{k}_r^{xy} \cdot \mathbf{r}_{xy}) +$$

$$(E_{rx}^0 E_{sx}^{0*} + E_{ry}^0 E_{sy}^{0*})E_{rx}^0 \exp[-i(2\mathbf{k}_r^{xy} - \mathbf{k}_s^{xy}) \cdot \mathbf{r}_{xy}] + (E_{rx}^{0*}E_{sx}^0 + E_{ry}^{0*}E_{sy}^0)E_{rx}^0 \exp(-i\mathbf{k}_s^{xy} \cdot \mathbf{r}_{xy}), \quad (S4)$$

where $E_{sx(y)}^0$ and $E_{rx(y)}^0$ are the complex *x(y)*-components of the signal and reference fields, respectively. It is seen that the third term, which is responsible for the reconstruction of the signal wave, contains the cross-term introducing the polarization crosstalk in the reconstruction of the *x*-component of the signal field (a similar term is found also in the expression for the *y*-component of the scattered field). The influence of the polarization crosstalk may result, for example, in reference wave scattering into orthogonal to that of the signal wave polarization as illustrated in our work.

## 2. Vectorial scattering holography: using surface plasmon polaritons as reference waves

Let us now consider a special case of vectorial scattering holography with the reference wave being represented by a surface plasmon polariton (SPP) propagating along the surface plane (Figure S2). In general, one can follow the reconstruction using the same expression [Eq. (S3)] for the SPP being the reference wave. It would suffice to change the reference field wavevector projection $\mathbf{k}_r^{xy}$ (on the surface plane) to the SPP wavevector $\mathbf{k}_{spp}$. There are, however, found several important differences and matters to be considered as compared to the general case above. First, in addition to the SPP being the evanescent wave (i.e., not propagating away from the hologram plane), the phase-conjugated (second) term is also describing evanescent waves, because the corresponding wavevector magnitude is larger than that of a free propagating wave: $|2\mathbf{k}_{spp} - \mathbf{k}_s^{xy}| > k_{spp} > k_0$, where $k_0$ is the wavenumber in air. The only propagating (away from the hologram plane) wave would therefore be the reconstructed signal wave [Figure S2(b)]. Second, even if one would assume that both signal and reference waves have the same polarization (which they cannot, because an SPP wave features a transverse spin found only in evanescent waves [S5]), the faithful reconstruction of a signal wave is only possible if one can neglect the SPP attenuation (because of its absorption) and depletion (because of its scattering by hologram nanoparticles). The first factor might become very important when considering the reconstruction of optical waves in visible or near infra-red (as in recent experiments [S6]), where the SPP absorption is significant. In this case, one should be careful when calculating the intensity interference pattern and reconstructing the signal wave with the corresponding scattering hologram, because an SPP amplitude is exponentially decreasing in the SPP propagation direction due to absorption (ohmic) loss. In order to accurately take the SPP attenuation into account, let us assign the *x*-axis to the SPP propagation direction [Figure S2(a)] and express the SPP field in the following form:

$$\boldsymbol{E}_{spp} = \boldsymbol{E}_{spp}^0 \cdot \exp(-\alpha x)\exp(-ik_{spp}x) \;, \tag{S5}$$

where $\alpha = 1/2L_{spp}$, with $L_{\text{spp}}$ being the SPP propagation length (over which the SPP intensity decreases as $1/e$), and $k_{spp} = \frac{2\pi}{\lambda}N_{spp}$, with $N_{\text{spp}}$ being the SPP effective index at the operating wavelength $\lambda$. The scattered field, in this case, is represented by a similar to that in Eq. (S3) expression, but with extra terms accounting explicitly for the SPP decay:

$$\boldsymbol{E}_{sc} \sim I\boldsymbol{E}_{spp} = \cdots + \left(\boldsymbol{E}_{spp}^{0*} \cdot \boldsymbol{E}_s^0\right)\boldsymbol{E}_{spp}^0 \exp(-2\alpha x)\exp\left(-i\boldsymbol{k}_s^{xy} \cdot \boldsymbol{r}_{xy}\right), \tag{S6}$$

where, for simplicity, only the reconstruction (third) term is written down. It is thereby seen that the signal wave reconstruction is not possible even without considering vectorial field properties, because the spatial domain of the signal field is cut short along the *x*-axis, when being reconstructed, to a stripe with the width of $L_{spp} = 1/2\alpha$.

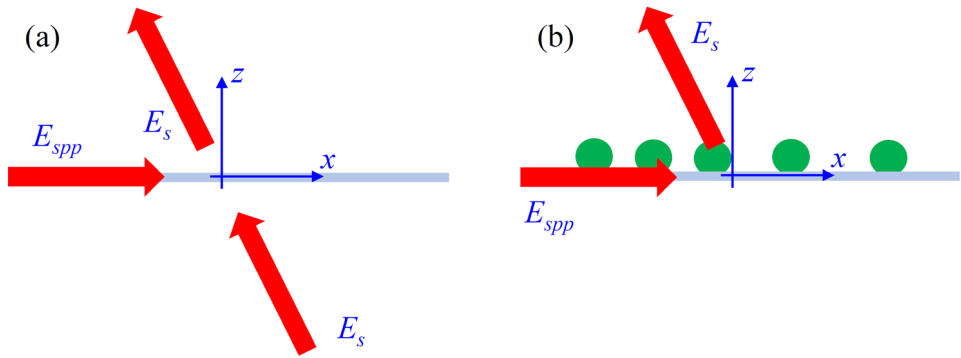

**Figure S2.** Schematic of vectorial scattering holography when using an SPP wave as the reference wave, illustrating (a) the determination of the intensity interference pattern produced by a signal, $E_s$, and reference SPP, $E_{\text{spp}}$, waves in the hologram *xy*-plane, and (b) the reconstruction of the signal wave upon interaction of the reference SPP wave with the arrays of surface nanoparticles.

In order, to avoid this narrowing down of signal spatial domain, one can simply use for calculating the intensity interference pattern the *artificially* growing reference SPP wave in the following form [cf. Eq. (S5)]:

$$\boldsymbol{E}_{a,spp} = \boldsymbol{E}_{spp}^0 \cdot \exp(\alpha x)\exp(-ik_{spp}x) \;. \tag{S7}$$

Considering the reconstruction of the signal wave with the *actual* SPP, which is exponentially decaying in the propagation direction, one obtains the corresponding (third) term as follows [cf. Eq. (S6)]:

$$\boldsymbol{E}_{sc} \sim \ldots + \left(\boldsymbol{E}_{spp}^{0*} \cdot \boldsymbol{E}_{s}^{0}\right) \boldsymbol{E}_{spp}^{0} \exp\left(-i \boldsymbol{k}_{s}^{xy} \cdot \boldsymbol{r}_{xy}\right) \,. \tag{S8}$$

In this case, the (relatively faithful) signal wave reconstruction is possible with respect to reconstructing the near-field amplitude distribution with the proper phase profile, ensuring the propagation in the specified (by the design) direction, although the problem with reconstructing polarization properties of the signal wave remains at the same level as discussed above. Note that the SPP depletion during the reconstruction, occurring due to the SPP scattering out of the surface plane by surface nanoparticles, can also be accounted for in a similar manner by simply adding the SPP attenuation by scattering to the SPP attenuation by absorption: $\alpha^* = \alpha + \alpha_{scat}$.

Finally, we consider the most relevant (to our present work) case of a cylindrically diverging SPP wave used as a reference wave that we treat in the same spirit as above. The cylindrically diverging SPP wave is thought to originate from a quantum emitter (QE) coupled to SPP waves (Figure S3).

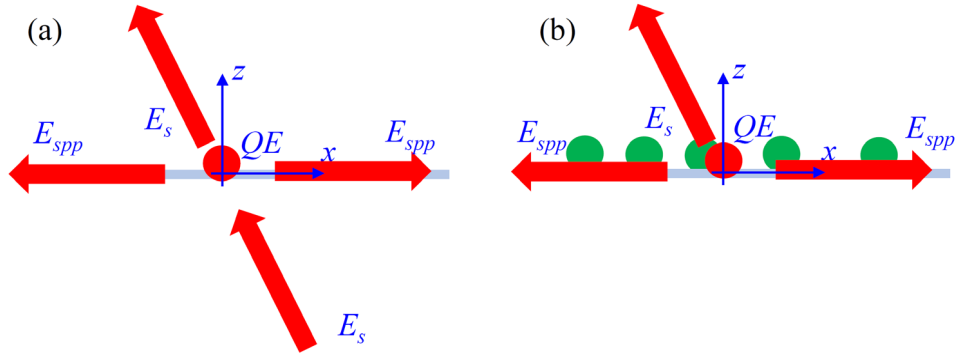

**Figure S3.** Schematic of vectorial scattering holography when using a radially diverging SPP wave as the reference wave, illustrating (a) the determination of the intensity interference pattern produced by a signal, $E_s$, and reference SPP, $E_{spp}$, waves in the hologram *xy*-plane, and (b) the reconstruction of the signal wave upon interaction of the reference SPP wave, which is originating from a quantum emitter (QE) coupled to SPP waves, with the arrays of surface nanoparticles.

Following the same line of reasoning, we should first remark that this SPP wave can be represented as follows [cf. Eq. (S5)] with the amplitude decrease due to the radial divergence explicitly incorporated:

$$\boldsymbol{E}_{spp} = \boldsymbol{E}_{spp}^{0} \cdot \left(1/\sqrt{r_{xy}}\right) \cdot \exp(-\alpha r_{xy}) \exp(-i \boldsymbol{k}_{spp} \cdot \boldsymbol{r}_{xy}) \,. \tag{S9}$$

Here the SPP field amplitude, $E_{spp}^0$, contains a radially oriented in-plane field component: $E_{spp,xy}^0 \sim (\cos\varphi, \sin\varphi)$, with $\varphi$ being the angle between the SPP propagation direction and the $x$-axis. If one uses this expression for determining the interference pattern, the signal wave reconstruction becomes problematic, because the spatial domain of the signal field would be limited to a small central area around the QE <u>irrespectively</u> the targeted signal spatial domain. In order, to avoid this shrinking of the signal spatial domain, one can use for calculating the intensity interference pattern the radially increasing reference SPP wave in the following form [cf. Eq. (S5)]:

$$\boldsymbol{E}_{a,spp} = \boldsymbol{E}_{spp}^0 \cdot \sqrt{r_{xy}} \cdot \exp(\alpha r_{xy}) \exp(-i\boldsymbol{k}_{spp} \cdot \boldsymbol{r}_{xy}) \ . \tag{S10}$$

In this case, the signal wave reconstruction would again be described by Eq. (S8), becoming thereby relatively faithful with respect to reconstructing the near-field amplitude distribution with the proper phase profile ensuring the propagation in the specified (by the design) direction.

### 3. Numerical simulation

Numerical simulations in this work are performed by commercial FDTD software package (FDTD Solutions, Lumerical Solutions). An electric dipole, viewed as an QE, is vertical to the substrate surface situated 50 nm above the top film. The wavelength in the simulation is $\lambda_0 = 670$ nm, which is consistent with the peak emission of the negative charge state of the NV centers in NDs [S7]. The permittivities of Ag and SiO$_2$ are taken from the handbook of optical constants of solids [S7], while the refractive index of HSQ is set as 1.41[S8]. First, we conduct the 2D FDTD simulation to estimate the effective index and the propagation length of SPP in the hybrid structures, which are used in the proposed holography approach to generate the interference patterns for scattering the QE-excited SPP waves to free space emission.

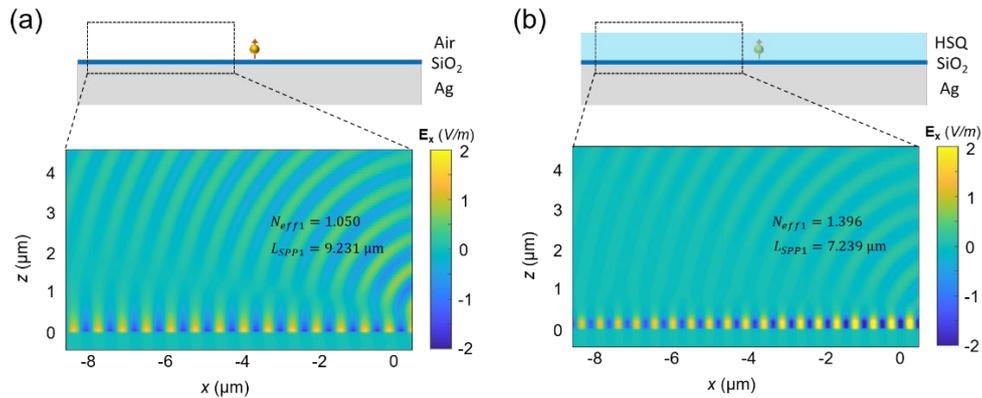

**Figure S4.** The effective indexes and propagation lengths of SPPs propagating along the interface between SiO$_2$ and Ag, (a) without HSQ layer (i.e., Air-SiO$_2$-Ag regime), (b) with HSQ layer (i.e., HSQ-SiO$_2$-Ag regime). The thickness of Ag substrate is 200 nm, followed by a 15 nm SiO$_2$ layer. The SPPs are excited by an electric dipole situated 50 nm above the SiO$_2$ layer surface and then propagate along the dielectric-metal interface. The dash line rectangles in the insets indicate the displayed areas in the simulation domain for presenting the distribution profiles of the electric field component **E**$_x$.

In Figure S4, we present the wavelength of SPPs, $\lambda_{spp}$, and effective indexes, $N_{eff}$, for both the scenarios without or with HSQ layer (180 nm thickness) on the top of substrate. The results show that for Air-SiO$_2$-Ag regime and HSQ-SiO$_2$-Ag regime the SPPs effective indexes $N_{eff1} = 1.050$ and $N_{eff2} = 1.396$, respectively. Correspondingly, the propagation lengths of SPP are $L_{SPP1} = 9.231\ \mu m$ and $L_{SPP1} = 7.239\ \mu m$. According to the effective medium theory, the effective index $N_{eff}^G$ of the hybrid structures can be weighted by the grating filling factor $\chi$ as $N_{eff}^G = (1 - \chi)N_{eff1} + \chi N_{eff2}$, and also the effective propagation length of SPP $L_{SPP} = (1 - \chi)L_{SPP1} + \chi L_{SPP2}$. For $\chi = 0.393$, we have $N_{eff}^G = 1.186$ and $L_{SPP} = 8\ \mu m$. It should be noted that these effective parameters, obtained by 2D simulation results, are only initial setting for designing hologram. The real filling ratio of the structure and its performance should rely on the 3D simulations. Actually, the small vibration of the pre-set filling factor, thereby changing $N_{eff}^G$ and $L_{SPP}$, and then the size of the structure and the domain, may somehow influence the beam divergence and the efficiency, but almost won't change the desired polarization and expected direction. It will be illustrated detaily in the following part.

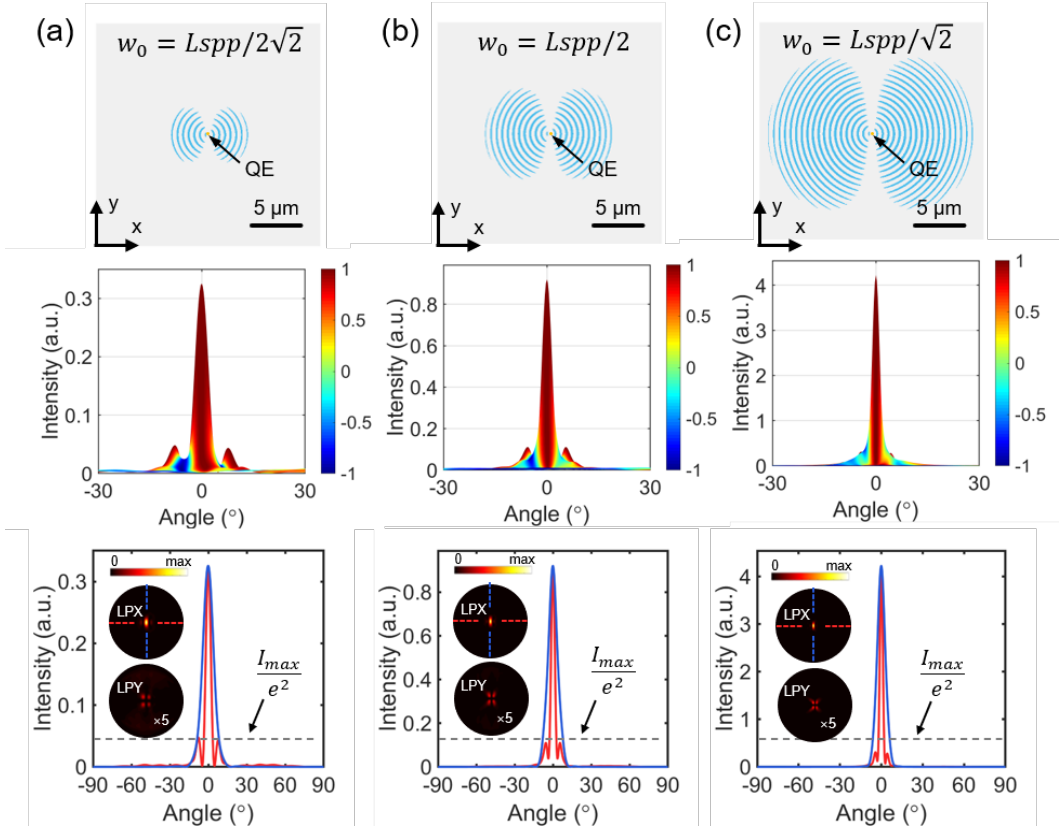

**Figure S5.** Directional linear polarization photon emission from hybrid SPP-QE coupled metasurfaces with (a) $w_0 = Lspp/2\sqrt{2}$, (b) $w_0 = Lspp/2$, and (c) $w_0 = Lspp/\sqrt{2}$. The first row is the top-view of the configurations. The second row is the superposition of the emission intensity and polarization, with the color representing the Stokes parameter $S_1$. The third row are the cross-section intensity curves of LPX, with the insets being the LPX and LPY emission patterns.

Figure S5 depicts the directional linear polarization photon emission from hybrid SPP-QE coupled metasurfaces with different Gaussian beam radius. As shown in the first row, the size of the metasurface domain increases with the Gaussian beam radius that is used to fit the SPP field. By overlapping the intensity of far-field photon emission and its polarization, it can be clearly seen that the emission peaks feature a very well linear polarization for all three cases, while the peaks become more and more shape for larger $w_0$. Moreover, by decomposing the LP polarization emission into LPX and LPY, it shows that the LPX dominates the polarization of emission photon, especially for larger $w_0$ (see the third row of Figure S5). The value of $I_{max}/e^2$ is used to value and mark the beam divergence. The corresponding beam divergences are show in the maintext [Figure 2(d)].

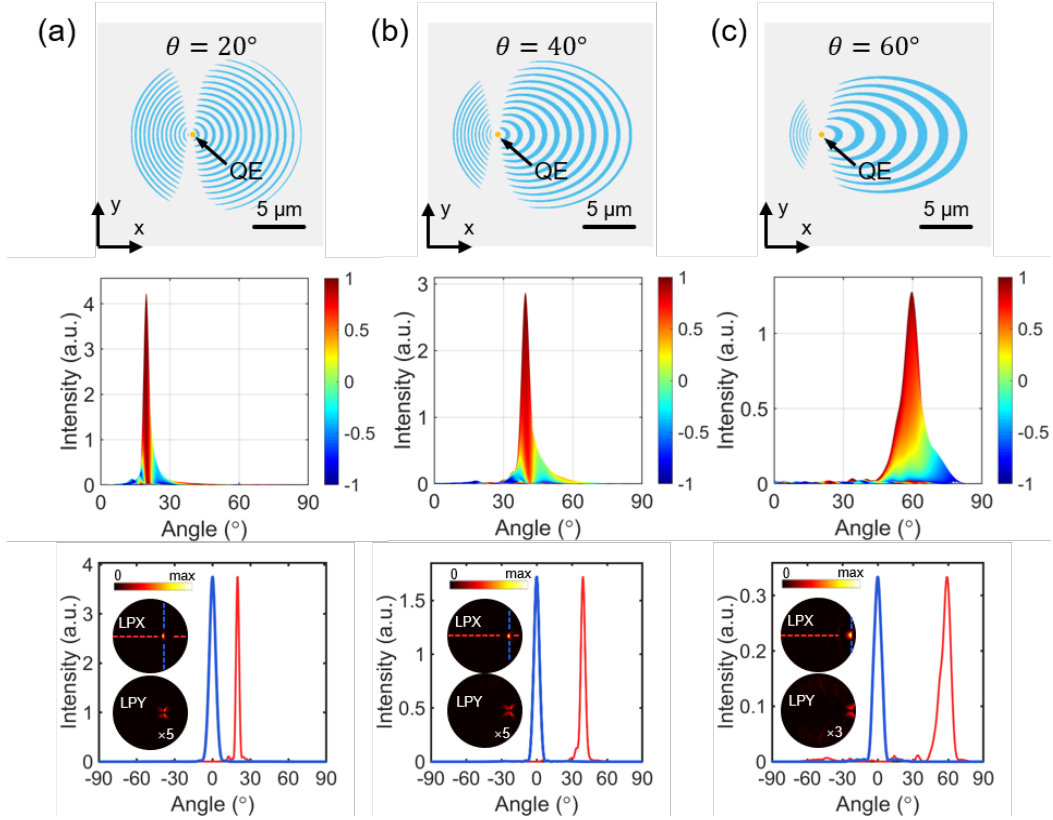

**Figure S6.** Off-axis linear polarization photon emission from hybrid SPP-QE coupled metasurfaces with (a) $\theta = 20°$, (b) $\theta = 40°$, and (c) $\theta = 60°$. The first row is the top-view of the configurations. The second row is the superposition of the polarization and emission intensity, with the color representing the Stokes parameter $S_1$. The third row are the cross-section intensity curves of LPX, with the insets being the LPX and LPY emission patterns.

Figure S6 presents the off-axis LP photon emission from hybrid SPP-QE coupled metasurfaces with $\theta = 20°$, $\theta = 40°$, and $\theta = 60°$. The first row is the top-view of the configurations. Interestingly, the structures are stretched in $x$ direction, with the period and width increasing in right side while decreasing in left side. Due to this characteristic, the number of rings of the generated pattern decreases with $\theta$, which results in the decrease of the outcoupling efficiency of SPP and thus a lower intensity of photon emission for $\theta = 60°$. In general, the photon emission features very well off-axis emission angle with desired linear polarization. It should be noted that the critical value, for judging whether to arrange a nanoelement in every pixel of the interference pattern, should also be very carefully chosen. However, it is hard (actually, also not so necessary) to evaluate the real filling ratio even after having the structure due to the nonuniform distribution of the scattering element in radial direction (for the circularly diverging SPP).

Similar to the process of LP photon emission from hybrid SPP-QE coupled metasurfaces, we investigate the off-axis radial polarization (RP) and right-hand circular polarization (RCP) photon emission. As shown in Figure S7, even for very large emission angel (e.g., 60°), the simulated results of the emission angle are in good agreement with the theory, demonstrating the effectivity and versatility of the proposed approach.

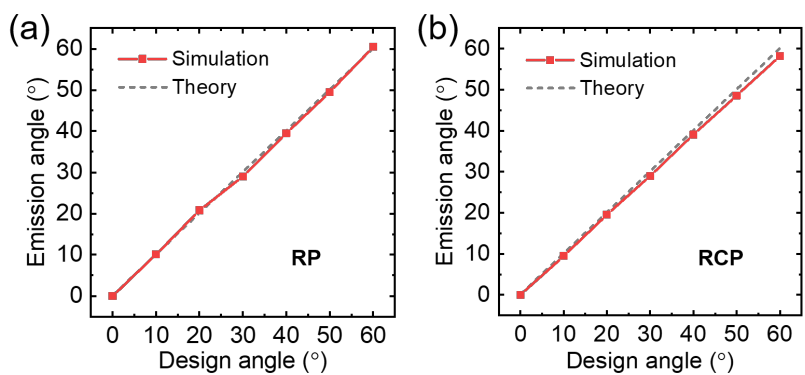

**Figure S7.** Off-axis (a) RP and (b) RCP photon emission from hybrid SPP-QE coupled metasurfaces with different designing emission angles.